\documentclass[
 reprint,
superscriptaddress,
 amsmath,amssymb,
 aps,
superscriptaddress,
 amsmath,amssymb,
 aps,longbibliography
]{revtex4-1}

\usepackage{graphicx}
\usepackage{dcolumn}
\usepackage{bm}
\usepackage{amsmath,amssymb,amstext,amsthm}
\usepackage{braket}
\usepackage{bbold}
\usepackage{bbm}
\usepackage{xcolor}
\usepackage{ gensymb }
\usepackage{soul}
\setstcolor{red}

\begin{document}

\preprint{APS/123-QED}
\setlength{\abovedisplayskip}{1pt}
\title{Psychophysical discrimination of radially varying polarization entoptic phenomena}

\author{D. A. Pushin}
\email{dmitry.pushin@uwaterloo.ca}
\affiliation{Institute for Quantum Computing, University of Waterloo,  Waterloo, ON, Canada, N2L3G1}
\affiliation{Department of Physics, University of Waterloo, Waterloo, ON, Canada, N2L3G1}
\affiliation{Centre for Eye and Vision Research, Hong Kong}
\author{C. Kapahi} 
\affiliation{Institute for Quantum Computing, University of Waterloo,  Waterloo, ON, Canada, N2L3G1}
\affiliation{Department of Physics, University of Waterloo, Waterloo, ON, Canada, N2L3G1}
\author{A. E. Silva} 
\affiliation{School of Optometry and Vision Science, University of Waterloo, Waterloo, ON, Canada, N2L3G1}
\author{D. G. Cory}
\affiliation{Institute for Quantum Computing, University of Waterloo,  Waterloo, ON, Canada, N2L3G1}
\affiliation{Department of Chemistry, University of Waterloo, Waterloo, ON, Canada, N2L3G1}
\author{M. Kulmaganbetov}
\affiliation{Centre for Eye and Vision Research, Hong Kong}
\author{M. Mungalsingh}
\affiliation{School of Optometry and Vision Science, University of Waterloo, Waterloo, ON, Canada, N2L3G1}
\author{T. Singh}
\affiliation{Centre for Eye and Vision Research, Hong Kong}
\author{B. Thompson}
\affiliation{Centre for Eye and Vision Research, Hong Kong}
\affiliation{School of Optometry and Vision Science, University of Waterloo, Waterloo, ON, Canada, N2L3G1}
\author{D. Sarenac}
\email{dsarenac@uwaterloo.ca}
\affiliation{Institute for Quantum Computing, University of Waterloo,  Waterloo, ON, Canada, N2L3G1}
\affiliation{Centre for Eye and Vision Research, Hong Kong}

\date{\today}


\pacs{Valid PACS appear here}


\begin{abstract}

The incorporation of structured light techniques into vision science has enabled more selective probes of polarization related entoptic phenomena. Diverse sets of stimuli have become accessible in which the spatially dependant optical properties can be rapidly controlled and manipulated. For example, past studies with human perception of polarization have dealt with stimuli that appear to vary azimuthally. This is mainly due to the constraint that the typically available degree of freedom to manipulate the phase shift of light rotates the perceived pattern around a person's point of fixation. Here we create a structured light stimulus that is perceived to vary purely along the radial direction and test discrimination sensitivity to inwards and outwards radial motion. This is accomplished by preparing a radial state coupled to an orbital angular momentum state that matches the orientation of the dichroic elements in the macula. The presented methods offering a new dimension of exploration serve as a direct compliment to previous studies and may provide new insights into characterizing macular pigment density profiles and assessing the health of the macula. 
\end{abstract}
\maketitle

\section{\label{sec:level1}Introduction}

The preparation, manipulation, and characterization of structured light ~\cite{rubinsztein2016roadmap,bliokh2023roadmap,chen2021engineering,ni2021multidimensional} have yielded impactful advances across a wide range of fields, including applications in high-resolution imaging, optical metrology, high-bandwidth communication, and material characterization~\cite{mair2001entanglement,Andersen2006,schwarz2020talbot,ritsch2017orbital, wang2012terabit, padgett2011tweezers,maurer2007tailoring,qplate,sarenac2018generation, milione20154, cameron2021remote,marrucci2011spin}. The backbone of structured light techniques is the tailoring of optical wavefronts and inducing a coupling between different degrees of freedom to obtain nontrivial propagation characteristics such as orbital angular momentum (OAM), non-diffraction, self-acceleration, and so on. Recent work has applied the structured light toolbox to the vision sciences~\cite{sarenac2020direct,sarenac2022human,gassab2023conditions,kapahi2023measuring}, enhancing the ability to explore entoptic phenomena through the use of polarization coupled OAM states.  Whereas uniformly polarized light induces the perception of a ``Haidinger brush'' which is a bowtie-like shape with two azimuthal fringes~\cite{haidinger1844ueber}, the structured light beams were shown to be capable of inducing a large and varying number of entoptic azimuthal fringes~\cite{sarenac2020direct}. A follow up study showed that the visual angle of the polarization related entoptic phenomena increases for a higher number of fringes~\cite{kapahi2023measuring}.

The human perception of polarization states of light is enabled by a series of radially symmetric dichroic elements (macular pigment) centred on the foveola in the human eye~\cite{misson2015human,misson2017spectral,misson2019computational,wang2022mathematical}. The spatial orientation of the radially oriented axonal fibers of retinal cells (mainly ganglion cell and photoreceptors) coupled with macular pigment effectively forms a weak radial polarizer within the retinal microstructure ~\cite{horvath2004polarized,mottes2022haidinger}. This feature has been exploited by polarized light studies of central visual field dysfunction and age-related macular degeneration~\cite{forster1954clinical,naylor1955measurement}, macular pigment density profiles~\cite{muller2016perception}, and the location of the fovea~\cite{vannasdale2009determination}. Early detection of age-related macular degeneration is of particular importance as it is a major cause of blindness worldwide~\cite{lim2012age}. 

Here we explore the use of structured light that induces the perception of fringes purely varying along the radial direction relative to the central point of vision. This is accomplished through the use of structured light that possesses a coupling between radial states and the OAM=2 orbital state that matches the symmetry of the dichroic macular pigment. The net result is the perception of entoptic radial fringes that move inwards/outwards relative to the center point. The macular pigment profile is azimuthally symmetric, sharply peaking at the center point of vision. Therefore, azimuthally varying entoptic motion is along the direction of constant macular pigment, while radially varying entoptic motion is along the direction with the most change in macular pigment. The presented method offers a new dimension of entoptic phenomena exploration and serves as a compliment to the previous studies that employed the use of azimuthally varying fringes.

\begin{figure*}
\centering\includegraphics[width=\linewidth]{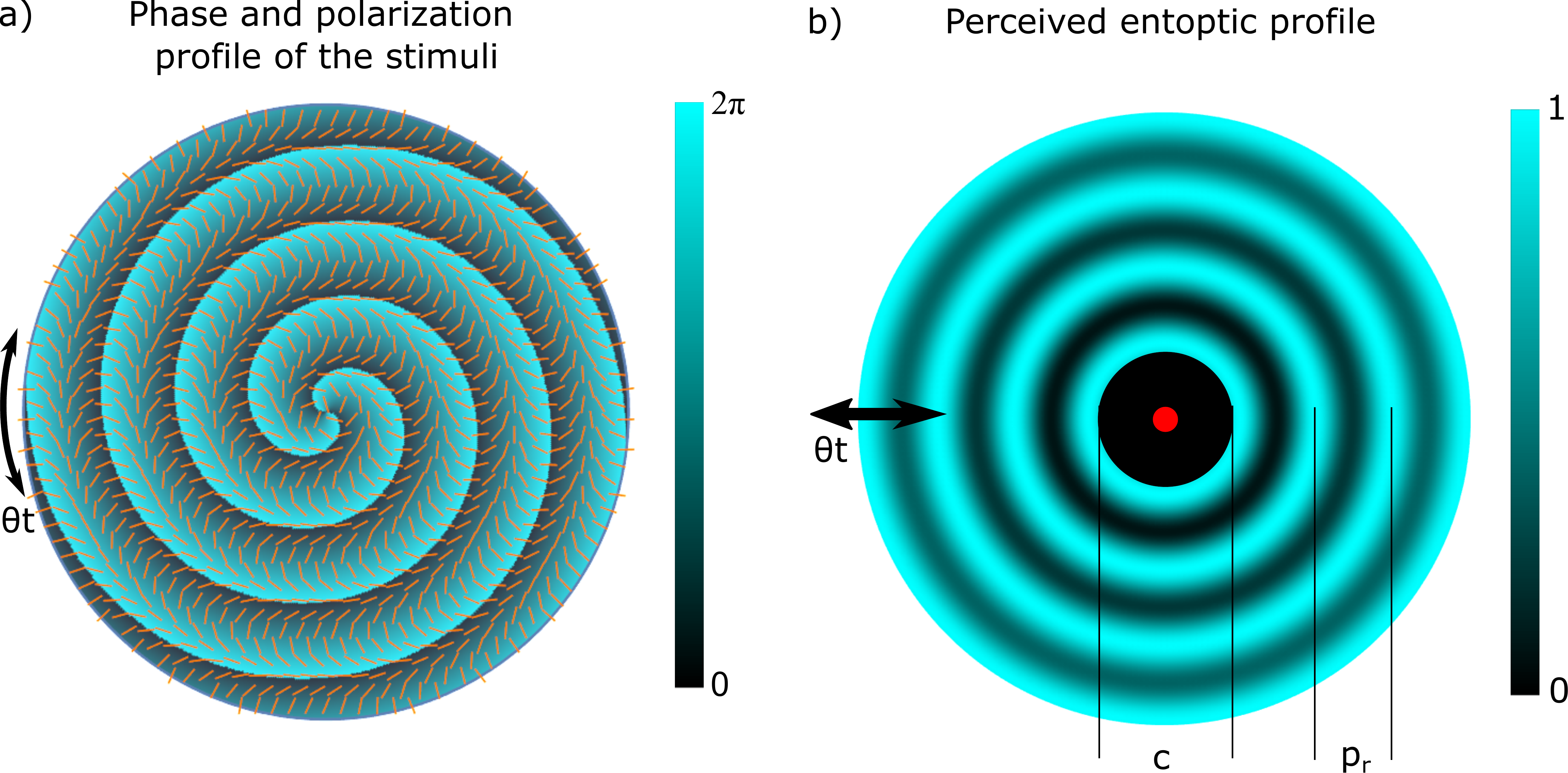}
\caption{a) The spatially dependant polarization (yellow bars) and phase profile (blue-black colormap) of the stimuli used in the study. b) Simulated entoptic pattern that would be perceived by a participant with a healthy macula. The eye's radial filter perceptually removes the spiral feature, and the final percept is that of radial waves. A radial blur is applied to depict the effect of decreasing macular pigment density along the radial direction. The clockwise/counterclockwise rotation of the beam in a) results in an inward/outward motion of the perceived radial pattern. The experimental stimulus included a central red light fixation guide, as shown. Finally, the size of the centrally blocked region, depicted in black, was varied to control the difficulty of the task and measure the peripheral extent of sensitivity to radial entoptic motion.
}
 \label{fig:fig1}
\end{figure*}

\section{\label{sec:level1}Methods}

The setup used in this study is described in detail in Ref~\cite{kapahi2023measuring}. Here we give a basic overview. The setup employs a spatial light modulator (SLM) to induce an arbitrary phase profile over the incoming beam. The pixel size of the SLM, and hence the available resolution, is 3 $\mu$m. A 4f imaging system in combination with a 20D Volk lens projects the profile from the SLM onto the participants retina, thereby removing diffraction and propagation effects. A small aperture illuminated by a red light is present in the middle of the stimulus to serve as a fixation point.

The target transverse wavefunction of the structured light beams for this study can be written as:

\begin{align}
	\ket{\Psi}=\frac{1}{\sqrt{2}}(1-\Pi[r/c])\left[e^{  i(n_r r+\ell\phi +\theta t)}\ket{R}
                +\ket{L}\right],
	\label{Eqn:Psi}
\end{align}

\noindent where $(r,\phi)$ are the transverse coordinates, $n_r$ and $\ell$ and the radial and OAM numbers, $\ket{L}$ and $\ket{R}$ are the right and left circularly polarized states, $\Pi[r/c]$ is the unit pulse function that sets the size ($c$) of the central obstruction area, and $\theta t$ is a time varying phase shift that dictates the speed of the perceived radial motion. A speed of $\theta = 1800 \degree/s$ was chosen in this study as this corresponds to a temporal frequency of $5 Hz$ at any spatial location, which allows for sufficient motion within a total presentation of 500 ms. The OAM value of $\ell=2$ has been chosen in order to match the symmetry of the dichroic macular pigment elements and induce a radial varying stimuli. The radial number $n_r=2\pi/p_r$ sets the period ($p_r$) of the radial gradients, and a radial period of $p_r=20 [px]$ on the SLM has been chosen as it roughly corresponds to 
 $p_r=0.9\degree$ of visual field. The spatially dependant phase and polarization profile of this state is depicted in Fig.~\ref{fig:fig1}a.

The entoptic profile that a participant will observe can be approximated by determining the intensity after Eq.~\ref{Eqn:Psi} passes through a radial polarization filter:

\begin{align}
	\text{I}=(1-\Pi[r/c])\cos^2\left(\frac{1}{2}(2\pi r/p_r+\theta t)\right),
	\label{Eqn:Intensity}
\end{align}

\noindent which is depicted in Fig.~\ref{fig:fig1}b. The figure also shows the red light fixation point and the expected radial blur which occurs with decreasing macular pigment along the radial direction. 

\begin{figure}
\centering\includegraphics[width=\linewidth]{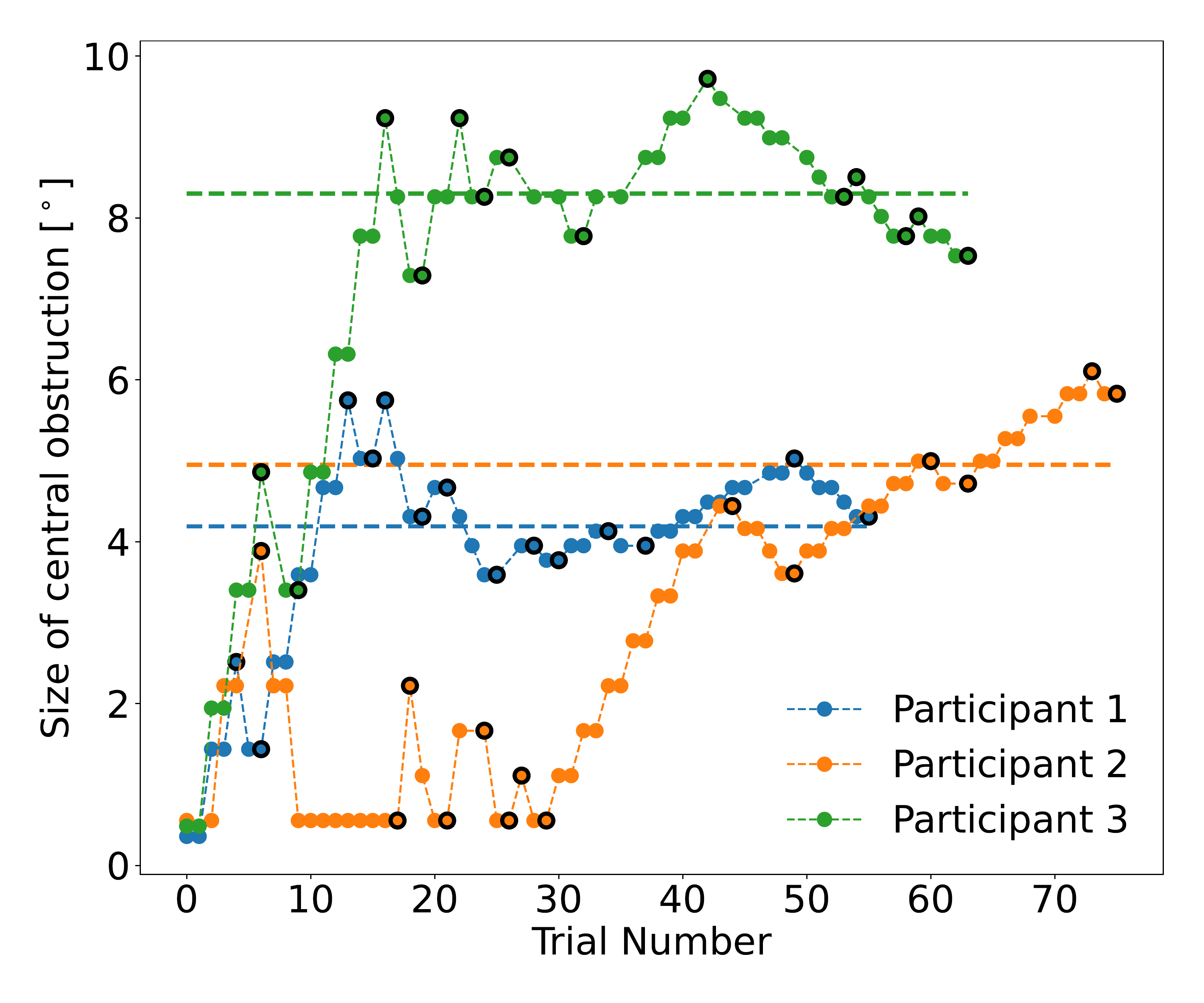}
\caption{The psychophysical study employed a staircase method where the size of the obstruction was varied according to the accuracy of the participant's responses. Shown are three different participants, with reversal points highlighted with a black outline. Threshold values from this 2-up/1-down staircase, computed from the final six reversal points are shown as dotted horizontal lines. The thresholds are converted into visual angle units through retinal imaging with the structured light stimulus, as described in Ref.~\cite{kapahi2023measuring}. 
}
 \label{fig:fig2}
\end{figure}

Eighteen participants were recruited to perform a motion direction discrimination task. All participants provided informed consent and were treated in accordance with the Declaration of Helsinki. All research procedures received approval from the University of Waterloo Office of Research Ethics. Before the main task, all participants performed a familiarization task in which an obstruction with a size of 20 pixels was presented 10 times. During these presentations, the stimulus duration was self-timed so that participants could learn to perform the task - most participants answered within a 2-5 second presentation window. All participants achieved at least 70$\%$ discrimination accuracy in the familiarization task.

Participants discriminated the radial motion of the entoptic pattern, either appearing to move inwards or outwards. The obstruction region ($c$) on the SLM was varied according to a 2-up, 1-down psychophysical staircase. The obstruction region became larger after two consecutive correct responses and became smaller after every incorrect response. This procedure resulted in an obstruction size threshold indicating the eccentric extent of sufficient polarization sensitivity to achieve at least 70.7$\%$ performance accuracy. A larger threshold indicates a larger eccentric range of polarization sensitivity.

A ``reversal'' was counted when the obstruction region had enlarged in previous trials but became smaller in a subsequent trial or vice verse. The task was completed after 14 reversals or after 90 total stimulus presentations. The arithmetic mean of the final six reversals was taken as the performance threshold. If the participant completed 90 stimulus presentations, then the final point was counted as a reversal. The initial obstruction size had a radius of 10 pixels and obstructed nothing beyond what the red fixation light obstructed. The obstruction radius was updated by 30 pixels after the first three reversals, by 20 after the following three reversals, by 10 after another three reversals, and finally by 5 for the remaining reversals. 

\begin{figure}
\centering\includegraphics[width=\linewidth]{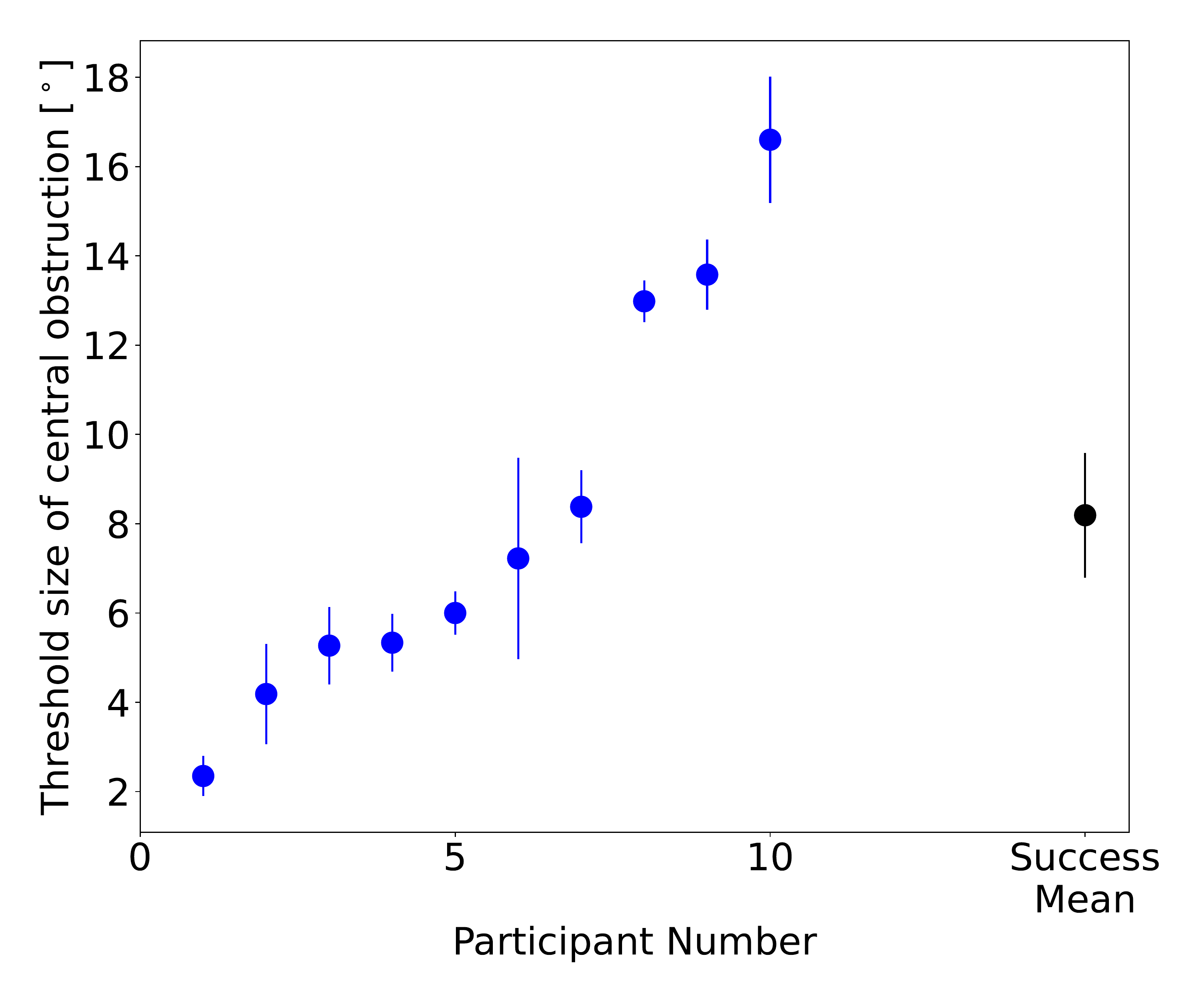}
\caption{Threshold values, plotted in ascending order, for each participant. The measured mean was $8.2\degree \pm 1.4\degree$ (standard deviation: 6.4) with a 95\% confidence interval of [$3.8^\circ$, $12.6^\circ$]. It was found that participant's perception of radial entoptic motion was noticeably harder than azimuthally varying motion.
$7/18$ participants exhibited 3 reversals at the performance floor, indicating poor performance. Furthermore, the mean threshold size indicates that at a comparable eccentricity roughly three times as much entoptic oscillation density was required to detect radial entoptic motion (this task) compared to azimuthal motion of Ref.~\cite{kapahi2023measuring}.}
 \label{fig:fig3}
\end{figure}


In the last step of the study, two retinal images were taken, the first with a standard fundus imaging system and the second with the structured light imaging system, allowing the psycophysically-defined threshold to be converted into degrees of visual angle~\cite{kapahi2023measuring}.

\section{\label{sec:level1}Study Results and Discussion}

The average obstruction size threshold (in degrees of visual angle) for the ten participants with successful staircase convergence was $8.2\degree \pm 1.40\degree$ (standard deviation: 6.4) with a 95\% confidence interval of [$3.8^\circ$, $12.6^\circ$]. This demonstrates that while polarization sensitivity varied between observers, it could be reliably measured even 8 degrees away from fixation. 
See Fig. 3 for subject-by-subject and summary results.

Staircases which contain more than 3 reversal points at the minimum central obstruction are considered ``failed'' and without sufficient sensitivity to measure a reliable threshold. It was found that a majority of participants had relatively more difficulty in reliably perceiving the radial motion when compared to azimuthal motion, and seven participants failed the task. Furthermore, one outlier participant, as defined by a threshold 3 standard deviations from the group mean, was excluded from analysis on the assumption of confounding eye movements. A possible cause of the relative difficulty could be that in comparison to azimuthally varying stimuli where the fringe oscillations are along the direction of constant macular pigment, radially varying entoptic motion is along the direction with the most change in macular pigment. Another  possible cause could be that the feature size of the azimuthally varying stimuli naturally scales for eccentricity, potentially compensating for decreased visual acuity outside of central fixation.

To compare the perception sensitivity of azimuthal and radial stimuli we can consider the relative size on the retina that one oscillation period occupies in each case. For entoptic motion that is a distance of $c/2$ away from the central point of fixation, the oscillation period is $\pi c/N$ for azimuthal stimuli while for radial stimuli the oscillation period is equal to $p_r$ and it is independent of $c$. Ref~\cite{kapahi2023measuring} considered a stimuli with N=11 azimuthal oscillations and reported an average central obstruction threshold of $c\approx9.5\degree$ which corresponds to an azimuthal oscillation period of $\approx 2.7\degree$ near the obstruction edge. Whereas the stimuli in this work possessed a three times smaller radial oscillation period of $p_r\approx 0.9\degree$ and we reported a comparable average central obstruction threshold of $8.2\degree$.  


The techniques described here can provide a complimentary set of data when analyzing a person's ability to perceive entoptic stimuli. The addition of the new degree of testing will allow us to perform more accurate characterization and  reconstruction of an individual's macular structure that is responsible for the perception of polarization related entoptic phenomena.

\section*{Acknowledgements}

This work was supported by the Canadian Excellence Research Chairs (CERC) program, the Natural Sciences and Engineering Research Council of Canada (NSERC) grants [RGPIN$-2018-04989$], [RPIN$-05394$], [RGPAS$-477166$],  the Government of Canada’s New Frontiers in Research Fund (NFRF) [NFRFE$-2019-00446$], the Velux Stiftung Foundation [Grant 1188], the InnoHK initiative and the Hong Kong Special Administrative Region Government, and the Canada  First  Research  Excellence  Fund  (CFREF). D.A.P. is grateful to Sonja Franke-Arnold for fruitful discussions.

\bibliography{OAM}

\end{document}